\documentclass[nofootinbib,twocolumn]{revtex4}

\usepackage{amssymb}
\usepackage{amsmath}
\usepackage{epsfig}
\usepackage{color}
\usepackage{graphicx}

\newcommand{\y}{\mathbf{e}_y}
\newcommand{\x}{\mathbf{e}_x}
\newcommand{\z}{\mathbf{e}_z}

\def\De{{\rm De}}

\def\d{{\rm d}}

\newcommand{\sh}{\dot{\boldsymbol{\gamma}}}
\newcommand{\st}{\boldsymbol{\tau}}
\newcounter{eqn}

\makeatletter
\newcommand{\putindeepbox}[2][0.7\baselineskip]{{%
    \setbox0=\hbox{#2}%
    \setbox0=\vbox{\noindent\hsize=\wd0\unhbox0}
    \@tempdima=\dp0
    \advance\@tempdima by \ht0
    \advance\@tempdima by -#1\relax
    \dp0=\@tempdima
    \ht0=#1\relax
    \box0
}}
\makeatother
\makeatletter
\let\start@align@nopar\start@align
\let\start@gather@nopar\start@gather
\let\start@multline@nopar\start@multline
\long\def\start@align{\par\start@align@nopar}
\long\def\start@gather{\par\start@gather@nopar}
\long\def\start@multline{\par\start@multline@nopar}
\makeatother

\begin{document}
%\setpagewiselinenumbers

\title{Small-amplitude swimmers can self-propel faster in viscoelastic fluids}
\author{Emily E. Riley}
\author{Eric Lauga}
\email{e.lauga@damtp.cam.ac.uk}

\affiliation{Department of Applied Mathematics and Theoretical Physics, 
University of Cambridge, CB3 0WA, United Kingdom.}
\date{\today}

\begin{abstract}
Many small organisms self-propel in viscous fluids using travelling wave-like deformation of their bodies or appendages. Examples include small 
nematodes moving through soil using whole-body undulations or spermatozoa swimming through mucus using flagellar waves. When self-propulsion occurs in 
a non-Newtonian fluid, one fundamental question is whether locomotion will occur faster or slower than in a Newtonian environment. Here we consider the 
general problem of swimming using small-amplitude periodic waves in a viscoelastic fluid described by the classical Oldroyd-B constitutive relationship. 
Using Taylor's swimming sheet model, we show that if all travelling waves move in the same direction, the locomotion speed of the organism is systematically 
decreased. However, if we allow waves to travel in two opposite directions, we show that this can lead to enhancement of the swimming speed, which is
physically interpreted as due to asymmetric viscoelastic damping of waves with different frequencies. 
A change  of the swimming direction is also possible. By analysing in detail the cases of swimming using two or three travelling waves, 
we demonstrate that swimming can be enhanced in a viscoelastic fluid for all Deborah numbers below a critical value or, for three waves or 
more, only for a finite, non-zero range of Deborah numbers, in which case a finite amount of elasticity in the fluid is required to increase the swimming speed. 
\end{abstract}

\maketitle

\section{Introduction}

A variety of small prokaryotic and eukaryotic organisms  exploit viscous forces from a surrounding fluid in order to self propel. The low  Reynolds number at which they swim  means there are no inertial effects, 
and work must be constantly expanded by the cells to produce  motion. This is achieved, for example, by the rotation of rigid helical appendages \cite{Purcell1997} or by the propagation of planar  travelling  waves 
along a flexible  flagellum \cite{Winet1977}. Our fundamental understanding of swimming cells  has increased dramatically with the advancement of imaging  techniques and computer power for more realistic numerical 
simulations~\cite{Powers2009}.

The vast majority of work on swimming at low Reynolds number has focused on swimmers moving in Newtonian fluids. However,  \emph{in vivo}, many self-propelled organisms  progress  through non-Newtonian 
fluids. Examples include the motion of cilia in lung mucus \cite{Liron1988}, nematodes travelling  though soil \cite{Wallace1967}, bacteria in their host's  tissue \cite{Suerbaum2002},  and spermatozoa 
swimming though cervical mammalian mucus \cite{Pacey2006}.  An important question is how a transition 
from  a Newtonian to a  non-Newtonian fluid affects the dynamics and kinematics of  micro-swimmers. In this paper, we use a simplified modelling approach to quantify   whether non-Newtonian stresses can help the micro-swimmers 
go faster or if they hinder their motion, and how  this affects their mechanical efficiency. 

Experimental studies have not yet reached a  clear consensus on whether  viscoelasticity increases or decreases swimming velocities. Instead a range  of results has been reported for different kinematics and rheological 
properties. Nematodes swimming in concentrated solutions of rod-like polymers  undergo an increase in swimming speed \cite{Arratia2013}. In that case, the  polymers,  aligned by the  stress caused by the nematode, form 
local nematic structures which give rise to shear-thinning  and aid the forward propulsion of the nematode. In contrast,  solutions of long flexible polymers with no shear-thinning  but strong elasticity lead to a  
decrease of the nematode's swimming speed \cite{Arratia2011}. An experiment imitating Taylor's classic swimming sheet \cite{Taylor1951} in rotational (planar) geometry shows exactly opposite effects with an increased locomotion
 in a Boger (constant-viscosity, elastic) fluid but a decrease  in a shear-thinning fluid \cite{Kudrolli2013}. Recently, the locomotion of flexible-tailed swimmers  was also shown to be enhanced  in a Boger fluid~\cite{Zenit2013}. 

Previous theoretical studies addressing motion in complex fluids  have considered a variety of kinematics, including undulatory motion \cite{Lauga2007,Loghin2013}, helical rotation \cite{Wolgemuth2007},   squirming \cite{Brandt2011,laipof1}, 
three-sphere models~\cite{Gaffney2013}, and paddlers~\cite{Loghin2013}. Methods that are ineffectual in a Newtonian fluid due to reversibility \cite{Purcell1977}, such as  flapping \cite{Lauga2008} or solid body  rotation \cite{pak12}, 
can also be exploited in a non-Newtonian setting to induce propulsion \cite{lauga_life,Lauga2011}. In the case of locomotion using helical flagella, small-amplitude helices always go slower, but  for larger amplitudes, a modest increase 
is possible \cite{liu2011}. In this paper, we focus on planar wave motion, a situation for which there is a wealth of  work starting with  Taylor's swimming sheet \cite{Taylor1951}.  
In the presence of a surrounding elastic  structure, non-Newtonian stresses were shown computationally to lead to faster and more efficient  swimming  \cite{Shelley2013}.  
Numerical simulations also demonstrated that for high-amplitude motion,  both shear-thinning  \cite{Loghin2013} and  polymeric Oldroyd-B fluids \cite{Shelley2010,spagnolie2013, Thomases2014} could lead to faster locomotion. 
{In particular, using simulations on finite swimming sheets it has been shown that front-back stress asymmetry together with swimmer flexibility leads to increased swimming speeds \cite{Thomases2014}.}

Analytical work on locomotion by waving  focuses on small-amplitude  motion. In the case of isolated swimmers, enhanced swimming was predicted theoretically to take place in gels \cite{Powers2010},  
Brinkmann fluids \cite{Leshansky2009} {and -- with the addition of elastohydrodynamic effects -- viscoelastic fluids \cite{Riley2014}}, but not in inelastic shear-thinning fluids \cite{velezcordero13}. 
Two nearby swimmers also synchronise faster in an elastic fluid than in a Newtonian medium  \cite{Elfring2010}. 
However, in the case of polymeric fluids, asymptotic  results predicted  a systematic  decrease of the swimming speed for all constitutive models, including all Oldroyd-like fluids \cite{Lauga2007} and general linear 
viscoelastic fluid models \cite{Powell1998} in the case of prescribed waveform swimming. A decrease  also takes place in the case of helical small-amplitude motion  \cite{Wolgemuth2007,FuWolgemuthPowers2009}. 
Provided {the prescribed} waving amplitude is small compared to its wavelength, it appears thus that an isolated swimmer is always slowed down by viscoelastic stresses.

In this paper, we consider mathematically the most general problem for planar locomotion using small-amplitude waves periodic both in space and in time. Specifically, we prescribe the shape deformation as a sum of waves travelling  
with different wavenumbers and  frequencies and in different directions and consider the resulting locomotion in a viscoelastic, Oldroyd-B fluid. We show that swimming in a non-Newtonian fluid at  small amplitudes need not 
always lead to slower swimming compared to the Newtonian case, provided the right combination of waves are considered. For swimming enhancement to be observed, different waves need to travel in  opposite directions, and the 
enhancement in that case results from the  asymmetric viscoelastic damping of waves with different frequencies. A change  of the swimming direction is also possible. After presenting the general derivations, and introducing 
a sufficient condition for enhanced locomotion, we analyse  in detail the cases of  two or three travelling  waves. The enhancement  in a viscoelastic fluid can be  obtained for all Deborah numbers below a critical value or, 
in the case of three waves or more, only if a finite amount of elasticity  is present in the fluid.

\section{General small-amplitude wave in a viscoelastic Fluid}
\subsection{Setup}
Analogous to Taylor's classic swimming calculation \cite{Taylor1951,Lauga2007}, an infinite inextensible sheet of negligible thickness is placed in a fluid and undergoes waving motion. The waveform of the sheet is prescribed, and results in swimming.  In the frame of the swimmer the oscillation of the  vertical position, $y(x,t)$,  of the sheet is described by
\begin{equation}
 y(x,t)=b\sum\limits_{n=-\infty}^{+\infty}\sum\limits_{m=-\infty}^{+\infty}\alpha_{n,m}e^{i(mkx-n\omega t)} ,
\label{gen}
\end{equation}
where  $x$ denotes the coordinate along the average sheet axis and $t$ time. In Eq.~\eqref{gen} the modes $n=0$ and $m=0$ are omitted as there is no mean deformation in $x$ or in time.   
The fluid is assumed to be located above the sheet along the $y >0$ direction In Eq.~\eqref{gen}, $b$ is the sheet amplitude, $k$ the fundamental wavelength and $\omega$ the fundamental frequency. 
We allow both positive and negative values of  the mode number $(m,n)$ in order to include waves travelling  in both directions along the sheet. 
The order-one complex coefficients $\alpha_{n,m}$ represent  dimensionless Fourier amplitude of each $(m,n)$ mode and since  $y$ is real they satisfy $\alpha_{-n,-m}=\alpha_{n,m}^*$. 
To simplify notation all sums over $n$ and $m$ from $-\infty$ to $+\infty$ will be denoted with a single summation symbol, $\sum_{n,m}$.

Upon non-dimensionalising $x$ by $k^{-1}$ and  $t$ by  $\omega^{-1}$, Eq.~\eqref{gen} becomes
\begin{equation}
 y(x,t)=\epsilon\sum\limits_{n,m}\alpha_{n,m}e^{i(mx-nt)},
\end{equation}
with a prefactor $\epsilon=bk$ defined as the ratio  of the sheet amplitude to its wavelength. We  assume that this ratio is small in this paper,  $\epsilon\ll1$, allowing  the swimming speed to be computed as an asymptotic expansion in $\epsilon$. 

As the sheet is infinite along the $z$  direction we can reduce the three-dimensional swimming problem to two dimensions. 
The velocity field is written as $\mathbf{u}=u_x\x+u_y\y$. This allows a streamfunction, $\psi(x,y,t)$, to be defined  such that  $u_x=\partial\psi/\partial y$ and $u_y=-\partial\psi/\partial x$, ensuring that  the flow remains incompressible. 

In order to find the streamfunction we  must first consider the boundary conditions imposed on the flow. On the waving sheet the no slip boundary condition enforces the velocity of the fluid at the sheet location to be the same as the velocity of the sheet, so that
\begin{equation}
\nabla\psi|_{x,y(x,t)}=\epsilon\sum\limits_{n,m}in\alpha_{n,m}e^{i(mx-nt)}\x.
\end{equation}
Far away from the sheet we expect that the flow will be unaffected by the wavemotion. Hence in the frame of the swimmer, the far field velocity will be the speed of the swimmer, but in the opposite direction. So if the steady swimming of the sheet is denoted  $-U\x$ then we have the boundary condition
\begin{equation}\label{inf}
\nabla\psi|_{x, \infty} = U\y,
\end{equation}
where the value of $U$ is to be determined. 
\subsection{Constitutive relationship: Oldroyd-B fluid}
The swimmer is self-propelling in a fluid described by the Oldroyd-B constitutive relationship, modelling a dilute solution of infinitely extensible polymers in a Newtonian solute  as a homogeneous continuum~\cite{Oldroyd1950,Phanbook}. 
In this classical model,  the shear viscosity is constant but the polymer elasticity affects the flow, giving rise to normal stresses. This is a good model for the Boger fluids  used in many non-Newtonian micro-swimmer experiments~\cite{Phanbook}.
Furthermore, from Ref.~\cite{Lauga2007}, we expect that to second order in $\epsilon$,  our asymptotic results  will remain valid for a large class of constitutive relationships.

 If $p$ denotes the  pressure and  $\st$ the deviatoric stress, Cauchy's equation of mechanical equilibrium in the absence of inertia is simply written
\begin{equation}
\label{cauchy}
\nabla p = \bf\nabla\cdot \st.
\end{equation}
In an Oldroyd-B fluid, the total deviatoric stress, $\st$, a combination of stresses from the Newtonian solvent $\st_s$, and those from the polymers $\st_p$, is written as $\st=\st_s+\st_p$. 
If $\eta_s$ denotes the solvent viscosity  and assuming that $\st_p$ follows a first-order Maxwell constitutive equation with relaxation time $\lambda$, 
{elastic modulus $G$, and polymer viscosity $\eta_p=G/\lambda$,} the total stress obeys \cite{Phanbook}
\begin{equation}
\st+\lambda\stackrel{\triangledown}{\st}=\eta\sh+\eta_s\lambda\stackrel{\triangledown}{\sh},
\label{oldB}
\end{equation}
where  $\sh$ is the shear rate tensor, defined as $\sh=\nabla {\bf u} + \nabla {\bf u}^T $, and $\eta=\eta_s+\eta_p$ is the sum of the solvent and polymer viscosities.  In Eq.~\eqref{oldB}, the upper-convected derivative  defines the rate of change of the  tensor $\bf A$ while it translates and deforms with the fluid and is written as 
\def\a{{\bf A}}
\begin{equation}
\stackrel{\triangledown}{\a} =\frac{\partial\a}{\partial t}+\mathbf{u}\cdot\nabla\a
-(\nabla\mathbf{u}^T\cdot\a+\a\cdot\nabla\mathbf{u}) .
\end{equation}
Upon non-dimensionalising stresses by $\eta\omega$ and shear rates by $\omega$, Eq.~\eqref{oldB} becomes
\begin{equation}\label{Deeq}
\st+\De\stackrel{\triangledown}{\st} = \sh+\beta\De\stackrel{\triangledown}{\sh}, 
\end{equation}
{where $\beta=\eta_s/\eta\leq 1$, and $\De=\lambda\omega$ is the Deborah number that describes the relative importance of viscoelasticity by 
comparing the relaxation time to the timescale on which the fluid is perturbed, given by $1/\omega$,   where $\omega$ is the fundamental waving frequency.}

\subsection{Asymptotic solution}
Since we have $\epsilon\ll1$ we seek to find solutions to the stress, streamfunction and velocity in terms of perturbative expansion in $\epsilon$, such that
\begin{align}
&\psi=\epsilon\psi_1+\epsilon^2\psi_2+\dots,  \\
&\st=\epsilon\st_1+\epsilon^2\st_2+\dots, \\
&U=\epsilon^2U_{2NN}+\dots.
\end{align}
The swimming velocity is expected to be quadratic in $\epsilon$, and so we  focus on the first and second-order solutions (the subscript $NN$ is used as a reminder that the final result for the swimming speed will quantify non-Newtonian swimming).

\subsubsection{Solution at order $\epsilon$}
The leading-order constitutive equation is linear and given by
\begin{equation}\label{order1}
\st_1+\De\frac{\partial\st_1}{\partial t}=\sh_1+\beta\De\frac{\partial\sh_1}{\partial t}\cdot
\end{equation}
This can be reduced into a streamfunction equation by taking its   divergence,  combining with Eq.~\eqref{cauchy}, and taking the curl to eliminate the pressure, leaving
\begin{equation}
\left(1+\beta\De\frac{\partial}{\partial t}\right)\nabla^4\psi_1 =0.
\end{equation}
The post-transient solution to Eq.~\eqref{order1} is found using Fourier notation and solving the biharmonic equation analytically, leading to
\begin{equation}
\psi_1=\sum\limits_{n,m}\alpha_{n,m}\frac{n}{m}(1+|m|y)e^{-|m|y}e^{i(mx-nt)},
\label{psi1}
\end{equation}
where the first-order boundary conditions,
\begin{subequations}
\begin{equation}
\nabla\psi_1|_{x,0}= \sum\limits_{n,m}in\alpha_{n,m}e^{i(mx-nt)}\x,
\end{equation}
and
\begin{equation}
\nabla\psi_1|_{x,\infty}=\mathbf{0},
\end{equation}
\end{subequations}
are satisfied. Clearly, the first-order solution is the same as the Newtonian case, and as expected there is no swimming at this order.

\subsubsection{Solution at order $\epsilon^2$}
At order $\epsilon^2$,  the constitutive equation, Eq.~\eqref{Deeq},  is given by
\begin{align}
\left(1+\De\frac{\partial}{\partial t}\right)\st_2&-\left(1+\beta\De\frac{\partial}{\partial t}\right)\sh_2 =\notag\\
&\De(\nabla\mathbf{u}_1^T\cdot\st_1+\st_1\cdot\nabla\mathbf{u}_1-\mathbf{u}_1\cdot\nabla\st_1)\notag\\
&-\beta\De(\nabla\mathbf{u}^T_1\cdot\sh+\sh\cdot\nabla\mathbf{u}_1-\mathbf{u}\cdot\nabla\sh_1).
\label{2nd}
\end{align}

Using Fourier notation of the form 
\begin{equation}
\mathbf{A}=\sum\limits_{n,m}\tilde{\mathbf{a}}^{(n,m)}e^{-int},
\end{equation}
for any tensor, vector, or scalar, the first-order constitutive equation, Eq.~\eqref{order1}, gives access to the Fourier component of the first-order stress as
\begin{equation}\label{tau1}
\tilde{\st}_1^{(n,m)}=\frac{1-in\beta\De}{1-in\De}\tilde{\sh}_1^{(n,m)}.
\end{equation}
As we are interested in the time-averaged swimming, it is sufficient to focus on the time-averaged version of Eq.~\eqref{2nd}.  We then use Eq.~\eqref{tau1} to express the mean of Eq.~\eqref{2nd} using the Fourier modes of its right-hand-side, and obtain
\begin{align}
\langle\st_2\rangle-\langle\sh_2\rangle=&\sum\limits_{n,m}\frac{\De(1-\beta)}{1-in\De}\times\notag\\
&(\nabla \mathbf{u}_{1}^{T\ast}\cdot\sh+\sh\cdot\nabla\mathbf{u}_{1}^{\ast}- \mathbf{u}_{1}^{\ast}\cdot\nabla\sh)^{(n,m)}.
\label{avft}
\end{align}

With the first-order streamfunction whose Fourier component is
\begin{equation}
 \tilde{\psi}_1^{(n,m)}= \alpha_{n,m}\frac{n}{m}(1+|m|y)e^{-|m|y}e^{imx},
\end{equation}
we obtain the Fourier modes of the flow velocity,
\begin{equation}
 \tilde{\mathbf{u}}_1^{(n,m)}=\alpha_{n,m}\frac{n}{m}e^{-|m|y}e^{imx}\left(\begin{array}{c} -|m|^2y \\ -(1+|m|y)im \end{array}\right),
\end{equation}
the velocity gradient,
\begin{align}
 \nabla\tilde{\mathbf{u}}_1^{(n,m)}=&\,\alpha_{n,m}\frac{n}{m}e^{-|m|y}e^{imx}\times\notag\\
&\left(\begin{array}{cc} -im|m|^2y & m^2(1+|m|y)\\ |m|^3y-|m|^2 & im|m|^2y \end{array}\right),
\end{align}
and the shear stress tensor,
\begin{equation}
 \tilde{\sh}_1^{(n,m)}= \alpha_{n,m}\frac{n}{m}e^{-|m|y}e^{imx}\left(
\begin{array}{cc} -2im|m|^2y & 2|m|^3y \\  2|m|^3y & 2im|m|^2y \end{array}\right).
\end{equation}
The divergence and curl are then taken, as before, to obtain an explicit equation for the second-order streamfunction as
\begin{align}
&\frac{d^4 \langle\psi_2\rangle}{dy^4}=\sum\limits_{n,m}-|\alpha_{n,m}|^2\frac{n^2}{m^2}\frac{(\beta-1)\De}{1-in\De}
\times\notag\\
&\,\,\frac{\d^2}{\d y^2}\Big[e^{-2|m|y}\left(-4im|m|^4y+4im|m|^5y^2-2|m|^3im\right)\Big].
\end{align}
Integrating with respect to $y$ three times, this gives
\begin{align}
&\frac{\d\langle\psi_2\rangle}{\d y}=Ay^2 + By + C+\notag\\
&\sum\limits_{n,m}|\alpha_{n,m}|^2\frac{n^2}{m^2}
\frac{(\beta-1)\De}{1-in\De}e^{-2|m|y}(-2im|m|^4y^2 + im|m|^2).
\end{align}
Given the form of the boundary conditions at infinity, Eq.~\eqref{inf}, we obtain $A=B=0$ and $C$ is equal to the second-order swimming speed, hence $C=U_{2NN}$. Its value can  be found using the time-averaged second-order boundary condition,
\begin{equation}
\frac{\d\langle\psi_2\rangle}{\d y}\Big|_{x,0}=\sum\limits_{n,m}nm|\alpha_{n,m}|^2,
\label{psi2}
\end{equation}
leading to
\begin{equation}
U_{2NN}=\sum\limits_{n,m}nm|\alpha_{n,m}|^2\left(\frac{1-in\De\beta}{1-in\De}\right).
\label{I.C.}
\end{equation}
Rewriting Eq.~\eqref{I.C.} with sums in $n$ and $m$ running  from $1$ to $\infty$ only, and using that  $\alpha_{-n,-m}=\alpha_{n,m}^*$,  leads to a simplified expression for the final result as
\begin{equation}\label{final}
U_{2NN}=2\sum\limits_{n\geq1}\sum\limits_{m\geq1}nm(|\alpha_{n,m}|^2-|\alpha_{n,-m}|^2)\left(\frac{1+\beta n^2\De^2}{1+n^2\De^2}\right),
\end{equation}
where the opposite-sign contributions of  waves travelling  in the $+x$  and $-x$ direction are apparent.

%%%%%%%%%%%%%%%%
\section{A sufficient condition for enhanced swimming}\label{engen}

The result in Eq.~\eqref{final} gives the leading-order swimming speed of the  swimming sheet with the  most general shape deformation periodic in both $x$ and $t$. 
When there are no viscoelastic effects $\De=0$, and the Newtonian result is recovered. We denote the swimming speed $U_{2N}$ in that case. 

As can be seen in Eq.~\eqref{final}, it is the value of the (dimensionless) frequency $n$ that affects the non-Newtonian change of each mode, not the value of the (dimensionless) wavenumber $m$. In order to gain  insight into the conditions 
for swimming to be enhanced or slowed down by the presence of viscoelastic stresses, let us focus on the simple case where only the modes $|m|=|n|$ are present. The sheet deformation is written now as a linear superposition 
of travelling  waves
\begin{equation}
 y=\epsilon\sum\limits_{n\geq1}\alpha_{+n}e^{in(x-t)}+\alpha_{-n}e^{in(x+t)},
\end{equation}
where $\alpha_{+n}$ and $\alpha_{-n}$ describes the $n$th mode wave travelling  to the right ($x>0$) and left ($x<0$) respectively. Using Eq.~\eqref{final} this leads to non-Newtonian swimming with speed
\begin{equation}
U_{2NN}=\sum\limits_{n\geq1}
a_n \left(\frac{1 +n^2\beta\De^2}{1+n^2\De^2}\right),
\label{Unnsumn}
\end{equation}
and Newtonian swimming with speed,
\begin{equation}
U_{2N} =  \sum\limits_{n\geq1}a_n,
\label{Unsumn} 
\end{equation}
where, we have further simplified notation such that
\begin{equation}\label{a_n}
 a_n = 2n^2(|\alpha_{+n}|^2-|\alpha_{-n}|^2),
\end{equation}
describes the superposition of mode $n$ waves in both directions. 
{Clearly,   for both Newtonian and non-Newtonian cases,  the addition of backwards waves always reduces the absolute value of the swimming speed.
Let us now focus on   the relative change in speed when comparing  swimming between a Newtonian and a  non-Newtonian fluid.}

Using only inspection we cannot, a priori, define a range of Deborah number where we expect to see an increase in speed from the Newtonian to the non-Newtonian swimming (i.e.~$U_{2NN}/U_{2N}>1$). 
In order to look for further insight, we consider the infinite and zero Deborah number limits. At zero Deborah number, {where there are no elastic effects,} the  ratio of swimming speeds is equal to 1. In the limit of 
large Deborah numbers $\De \gg 1$, {where elastic effects dominate,} it is straightforward to get from Eq.~\eqref{Unnsumn} that $U_{2NN}/U_{2N}=\beta < 1$, and thus swimming is  always eventually decreased. As the value of $\De$ increases from zero to infinity, 
the speed ratio could monotonically decrease from $1$ to $\beta$, in which case no enhancement would be seen, or non-monotonically, where enhancement could take place. 

Our numerical simulations indicate that in the cases where the speed ratio does go above 1, then in most cases it is always increasing in  the neighbourhood of $\De=0$ before monotonically decreasing to $\beta$ (see numerical results in Fig.~\ref{speedratio} and discussion below). In order to characterise the behavior around $\De=0$,  we can compute derivatives and Taylor-expand the ratio of swimming speeds. The first derivative ${\partial U_{2NN}}/{\partial\De}$ evaluated at $\De=0$ is zero because the 
swimming speed depends quadratically on the Deborah number. However, the second derivative (the curvature) is non-zero, and is given by
\begin{equation}\label{curv}
 \frac{\partial^2U_{2NN}}{\partial\De^2}\bigg|_{\De=0} = \sum\limits_{n\geq1} 2n^2a_n(\beta-1).
\end{equation}
When it is divided by the Newtonian swimming speed, Eq.~\eqref{Unsumn}, the above
gives access to the curvature of $U_{2NN}/U_{2N}$ at $\De=0$ (this is equivalent to taking the first derivative of the speed ratio with respect to $\De^2$). 
 If that curvature is positive, then  faster swimming occurs in the neighbourhood of $\De=0$. As we always have $\beta<1$, the curvature is positive if there is a sign difference between the sums in 
 Eqs.~\eqref{Unsumn}-\eqref{curv} and therefore a sufficient condition for enhanced swimming is the kinematic condition 
\begin{equation}\label{cond}
\left[ \sum\limits_{n\geq1} a_n\right]\times \left[\sum\limits_{n\geq1} n^2a_n\right] < 0.
\end{equation}
In order to achieve the condition in Eq.~\eqref{cond}, waves travelling  in opposite directions are required. Indeed, for example if all  $a_{n}$ amplitudes are positive, then it is easy to see from Eq.~\eqref{Unnsumn} 
that each $a_{n}$ mode decreases in amplitude, resulting in an overall decrease in  magnitude of the speed. If there are waves  travelling  in both direction, i.e.~at least one $\alpha_{-n}\neq0$ and one 
$\alpha_{+n}\neq 0$, then they need different combinations  of amplitudes and frequencies in order to satisfy the condition in Eq.~\eqref{cond}. Hence a combination of positive and negative $a_n$ values are required.

Physically, the  increase in swimming speed between Newtonian and viscoelastic fluids seen here arises from the fact that the damping caused by a non-zero Deborah number affects modes with different frequencies differently. 
Specifically, the damping term of the  form $({1 +n^2\beta\De^2})/({1+n^2\De^2})$ decreases monotonically with $n$. Modes with higher frequencies are therefore damped more than those with lower values of $n$, 
which provides a mechanism for enhanced swimming. 

For illustration,  consider two waves travelling  in opposite directions with the high-frequency ($n$) wave travelling  along the $-x$ direction ($a_{n}<0$)  and the low-frequency ($m$) one along the $+x$ direction ($a_{m}>0$). Then their  respective 
amplitudes be such that the resulting Newtonian swimming speed is positive, $U_{2N}>0$. In the viscoelastic fluid, the $a_{n}$ wave will be damped more than the $a_{m}$ wave, as $n>m$. 
On one hand, decreasing the magnitude of the  $a_{n}$ wave will increase the swimming speed while on the other hand, decreasing the $a_{m}$ mode will hinder the swimming velocity -- it is thus a matter of relative decrease. If the  wave amplitudes are such that 
the gain found by  suppressing the $a_{n}$ wave more than compensates for the damping of the $a_{m}$ wave, then the non-Newtonian swimming speed will be above the Newtonian one,  $U_{2NN}>U_{2N}$. If the wave amplitudes are such 
that $U_{2N}<0$, then a similar reasoning might be used to lead to $U_{2NN}>0$ and in that case, viscoelasticity might lead to a reversal of the direction of locomotion.

%%%%%%%%%

\section{Superposition of two travelling  waves: continuous enhancement}
\label{twwvs}

We now consider in detail simple cases. We start by swimming using two travelling  waves, and show that in this case the sufficient condition described above is in fact necessary: when enhancement takes place, it will lead to faster swimming for all Deborah numbers below a critical value. 
In order to  analytically describe situations where faster  swimming can occur, two simple waveforms each containing two waves travelling  in opposite directions will be considered. Clearly these two travelling  waves must have different frequencies, otherwise they are both damped in the same proportion by viscoelasticity and the swimming speed decreases. 

\subsection{Superposition of two travelling  waves with identical wave speeds}
\label{ex}
\setcounter{eqn}{0}
\begin{figure*}[t]
\includegraphics[width=0.9\textwidth]{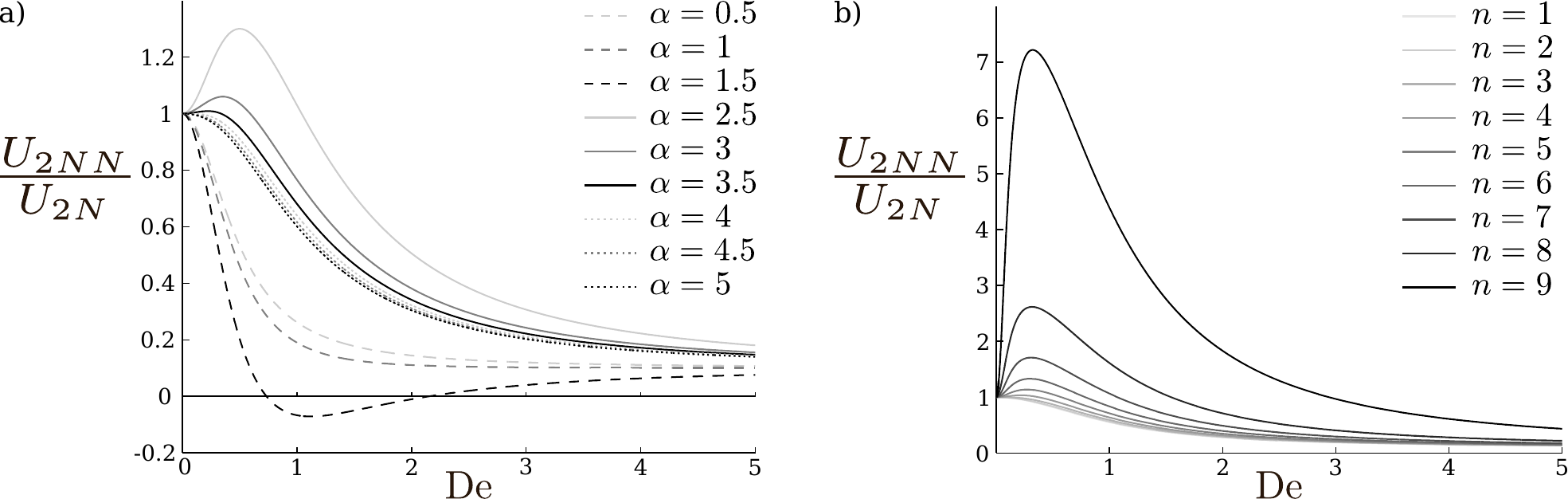}
\caption{Ratio between the non-Newtonian swimming speed $U_{2NN}$, and the Newtonian value $U_{2N}$, as a function of the Deborah number, $\De$, for various values of the relative wave amplitude $\alpha$,  and 
frequency ratio $n$, in the waveform from Eq.~\eqref{ss}. Here we have chosen $\beta=0.1$. 
Left: fixed value of $n=2$  and a range of $\alpha$ values (between 0.5 and 5). 
Right:  fixed value of $\alpha=9.5$ and $n$ ranging between $n=1$ and $9$.}
\label{speedratio}
\end{figure*}

An example of two waves with different  frequencies modes, amplitudes, and wave direction but identical magnitude of wave speed is given by 
\begin{equation}
 y(x,t)=\epsilon\left[\alpha\sin(x-t)+\sin n(x+t)\right],
\label{ss}
\end{equation}
where $\alpha$ is {the  dimensionless ratio of  amplitudes between the two waves.}
Using Eqs.~\eqref{Unnsumn} and ~\eqref{Unsumn} for the sinusoidal waveform in Eq.~\eqref{ss} we get the second-order Newtonian swimming speed as
\begin{equation}\label{33}
 U_{2N}=\frac{1}{2}(\alpha^2-n^2),
\end{equation}
while the second-order non-Newtonian swimming speed is given by
\begin{equation}\label{34}
 U_{2NN}=\frac{\alpha^2}{2}\bigg(\frac{1+\beta\De^2}{1+\De^2}\bigg)-\frac{n^2}{2}\bigg(\frac{1+n^2\beta\De^2}{1+n^2\De^2}\bigg).
\end{equation}
To find where faster swimming occurs, we compute as above the  second derivative of the swimming ratio, $U_{2NN}/U_{2N}$ with respect to $\De$ at 
  $\De=0$, giving
\begin{equation}\label{kappa}
\frac{\partial^2}{\partial \De^2}\left(\frac{U_{2NN}}{U_{2N}} \right)\bigg|_{\De=0}= 2(\beta-1)\bigg(\frac{\alpha^2-n^4}{\alpha^2-n^2}\bigg).
\end{equation}
This is positive (i.e.~upwards curving from $U_{2NN}/U_{2N}=1$) when $n<\alpha<n^2$.
Hence faster swimming requires the relative amplitude between the two waves to lie in a precise interval. If $\alpha$ is too small the behavior is dominated by the $-x$ wave while if it is too large the dynamics is 
dominated by the $+x$ wave.   At higher modes, the range of amplitudes available to the swimming sheet that can produce faster swimming in a non-Newtonian environment compared to a Newtonian one is increased.

\begin{figure*}[t]
\includegraphics[width=0.75\textwidth]{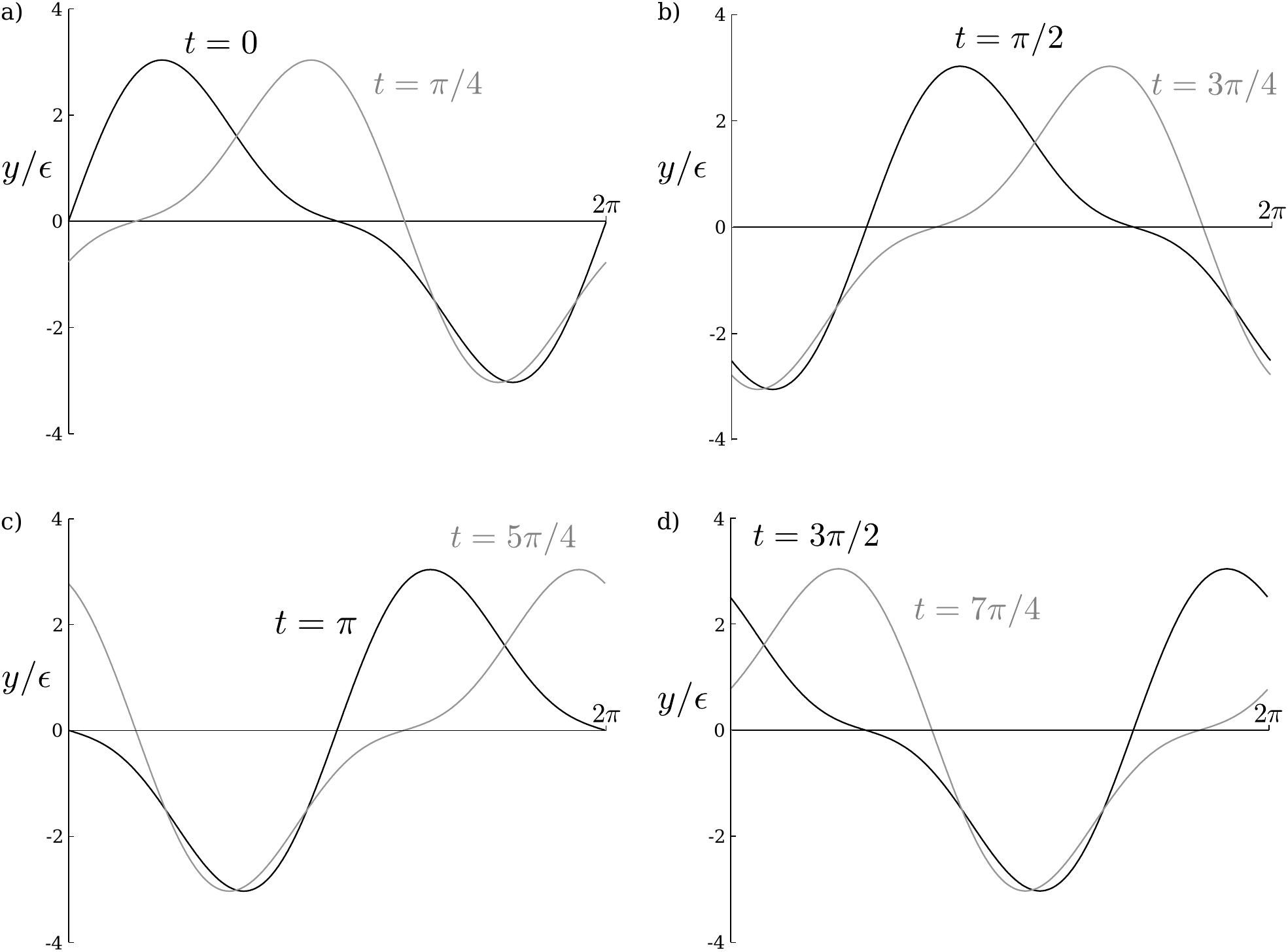}
\caption{Illustration of a waveform  producing faster swimming in a non-Newtonian fluid. 
The waveform is described by Eq.~\eqref{ss} with $\alpha=5/2$ and $n=2$, and corresponds to a swimming speed ratio as shown by the uppermost  solid grey line in Fig.~\ref{speedratio}a.  
The black lines in each of the four figures show the  waveform at dimensionless times $0$, $\pi/2$, $\pi$ and $3\pi/2$, respectively, and the grey  lines show the evolution of the wave an eighth of a period later, 
to show how the wave travels and changes shape during its period.}
\label{wave}
\end{figure*}

We illustrate in Fig.~\ref{speedratio} these results numerically. We plot the ratio of swimming velocities, $U_{2NN}/U_{2N}$, as a function of the Deborah number, $\De$, for a range of values of both $n$ and $\alpha$. 
We choose a fixed value of $\beta=0.1$.
The computational results confirm that when enhanced swimming is obtained, the speed ratio first increases in the neighbourhood of $\De=0$ before monotonically decreasing to $\beta$. This validates the curvature analysis 
as a proxy for predicting enhanced swimming, and indeed faster swimming in a non-Newtonian fluid is seen in the range $n<\alpha<n^2$.  An illustration of travelling  wave that swims faster in a non-Newtonian fluid is 
shown in Fig.~\ref{wave}, with $n=2$ and $\alpha=5/2$. This waveform corresponds to the speed ratio shown as the  uppermost  solid grey line in Fig.~\ref{speedratio} with a maximum of $U_{2NN}= 1.3$ at $\De= 0.5$.

\setcounter{eqn}{0}

Further analytical insight can be provided by noting from Fig.~\ref{speedratio} that the peak swimming speed ratio occurs when $\De$ is order one. Dividing the result in 
Eq.~\eqref{34} by that in Eq.~\eqref{33} and taking a first derivative with respect to $\De$ we can compute the value of the Deborah number at which the velocity ratio is extremized. It occurs for two values of $\De$ given by
\begin{equation}
 \De_{1\ast}=\sqrt{{\frac{n^2-\alpha}{n^2(\alpha-1)}}} \,\,\text{and}\,\, \De_{2\ast}=0.
\end{equation} 
For $\alpha$ above $n^2$ the only solution is the maximum value of 1 occurring at $\De_{2\ast}=0$. When $\alpha$ crosses below $n^2$  a maximum is created near  $\De_{1\ast}=0$, and increases as $\alpha$ decreases. When $\alpha=n$ a transition occurs where 
$\De_{1\ast}$ changes from a  maximum point ($n < \alpha$) to a minimum  ($\alpha<n$); its value at that point is  $\De_{1\ast}=1/n$. It remains a minimum until $\alpha$ crosses the value 1, below which the only solution is the maximum of 1  at $\De_{2\ast}=0$.

A final point of interest  in  Fig.~\ref{speedratio} is the fact, as discussed above, that the ratio between the swimming speeds can become negative. 
In these cases, the swimmer would then  swim in different directions in the Newtonian and non-Newtonian fluids, as was already noted in Ref.~\cite{Wolgemuth2007}. 
This occurs when $\alpha < n$, and the speed ratio goes through a  minimum before increasing back towards $\beta$ at large Deborah numbers. 
The reversal of swimming occurs when there is a difference in sign between Eq.~\eqref{33} and Eq.~\eqref{34}, which corresponds to the amplitude range
\begin{equation}
 \sqrt{\frac{n^2(1+\beta n^2\De^2)(1+\De^2)}{(1+n^2\De^2)(1+\beta\De^2)}} < \alpha < n.
\end{equation}
This result is reminiscent of a recent study on reciprocal (time-reversible) motion in a worm-like micellular solution, which showed that the direction and the speed of the 
swimmer could be changed when distinct Deborah numbers are reached~\cite{Gagnon2014}. Finally, we  can also find a range of a values for which the swimmer will not only change direction but will also swim with a larger magnitude, which  occurs when
\begin{equation}
  \sqrt{\frac{n^2(1+\beta n^2\De^2)(1+\De^2)}{(1+n^2\De^2)(1+\beta\De^2)}-1} < \alpha < n.
\end{equation}
Here the swimming speed ratio becomes negative and less than $-1$.

\subsection{Necessary vs.~sufficient condition for enhanced swimming}
The sufficient condition for enhancement derived in \S\ref{engen} detailed the conditions required for an  upwards curving  of the swimming speed ratio from zero Deborah number. 
In order to study if this sufficient condition is also necessary, we search analytically for the conditions leading to $U_{2NN}>U_{2N}$, leading to
\begin{equation}
 0<\De<\sqrt{\frac{n^4-\alpha^2}{n^2(\alpha^2-n^2)}}\equiv \De_a.
\end{equation}
This condition requires $n<\alpha<n^2$, and defines the range of Deborah number where forward swimming enhancement is achieved, namely $[0,\De_a]$. 
If we enforce the curvature to be negative then we cannot find a set of viable parameters for which $U_{2NN}>U_{2N}>0$, thus showing that the  sufficient condition is also  necessary  when two modes are considered: 
in the case of two waves, if forward swimming enhancement is ever to be obtained, it will take place for any Deborah number below a critical value $\De_a$.

\subsection{Swimming efficiency}
\setcounter{eqn}{0}
\begin{figure*}[t]
\includegraphics[width=0.8\textwidth]{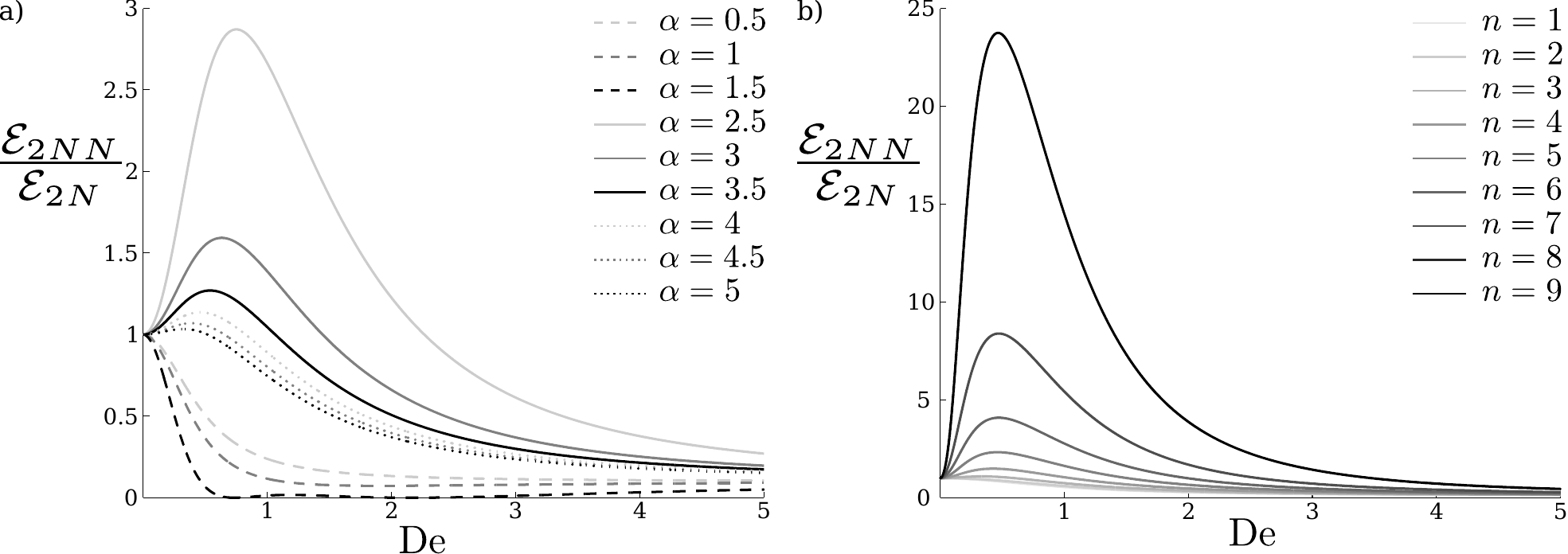}
\caption{Ratio of the swimming efficiency in a non-Newtonian fluid compared to its Newtonian counterpart as a function of $\De$: (a) $n=2$ for a range of  values of $\alpha$;  
(b) $\alpha=9.5$ and $1\leq n \leq 9$. The waveform is the one described in Eq.~\eqref{ss}.}
\label{effic}
\end{figure*}

We now turn to energetic considerations. The  rate of viscous dissipation in the fluid as the sheet is swimming is equal to the volume integral of $\st:\sh$ in the fluid. 
At leading order we therefore have to integrate $\st_1 : \sh_1$. With the  general waveform in Eq.~\eqref{gen}, the dimensional second-order dissipation rate in the non-Newtonian fluid per unit length in the $\z$ direction is easily found and we obtain $W = \epsilon^2W_{2NN}+\dots$ with
\begin{align}\label{workNN}
 W_{2NN} =\sum\limits_{n\geq1}\sum\limits_{m\geq1}&8\pi\eta\omega^2mn^2\notag\\
&\left(\frac{1+n^2\beta\De^2}{1+n^2\De^2}\right)\left(|\alpha_{n,m}|^2-|\alpha_{n,-m}|^2\right).
\end{align}
The result in Eq.~\eqref{workNN} should then be compared with its Newtonian counterpart. 

Let us consider for illustration the waveform in Eq.~\eqref{ss}. In that case, the ratio of the work done against the non-Newtonian fluid compared to the Newtonian one  is given by
\begin{align}
 \frac{W_{2NN}}{W_{2N}} =&\frac{\eta}{\eta_N(n^3+\alpha^2)}\times\notag\\
&\left[\alpha^2\left(\frac{1+\De^2\eta_{s}/\eta}{1+\De^2}\right)+n^3\left(\frac{1+n^2\De^2\eta_{s}/\eta}{1+n^2\De^2}\right)\right],
\end{align}
where $\eta_N$ is the Newtonian viscosity. In order to contrast the locomotion in the polymeric fluid with that in the solvent alone, we then take $\eta_N=\eta_s$. Furthermore, as is done traditionally, the swimming  efficiency  is defined as
\begin{equation}
\mathcal{E}= \frac{\eta U^2}{W}\cdot
\end{equation} 
In order to compare the efficiency of swimming in the different fluids, we compute the ratio
\begin{equation}
 \frac{\mathcal{E}_{2NN}}{\mathcal{E}_{2N}} = \frac{\eta U_{2NN}^2}{W_{2NN}}\frac{W_{2N}}{\eta_N U_{2N}^2}\cdot
\end{equation}
The ratio of the work and viscosity in the two different fluids, $\eta W_{2N}/\eta_N W_{2NN}$, is always greater than $1$ for non-zero Deborah number, meaning that  when  the swimming speed ratio $U_{2NN}/U_{2N}$ is greater than $1$, the swimming efficiency is automatically  always increased.  

We plot the ratio of efficiencies   against $\De$ for a range of relative wave amplitude $\alpha$, and wavenumber ratio $n$, in Fig.~\ref{effic}, where $\eta_s/\eta=\beta=0.1$.    Clearly, an increase in swimming speed is correlated with an increase in efficiency, but increased efficiencies can in fact be obtained without enhanced swimming. 
 Indeed, increased efficiency is obtained as soon as 
\begin{figure}[b]
\centerline{ \includegraphics[width=0.35\textwidth]{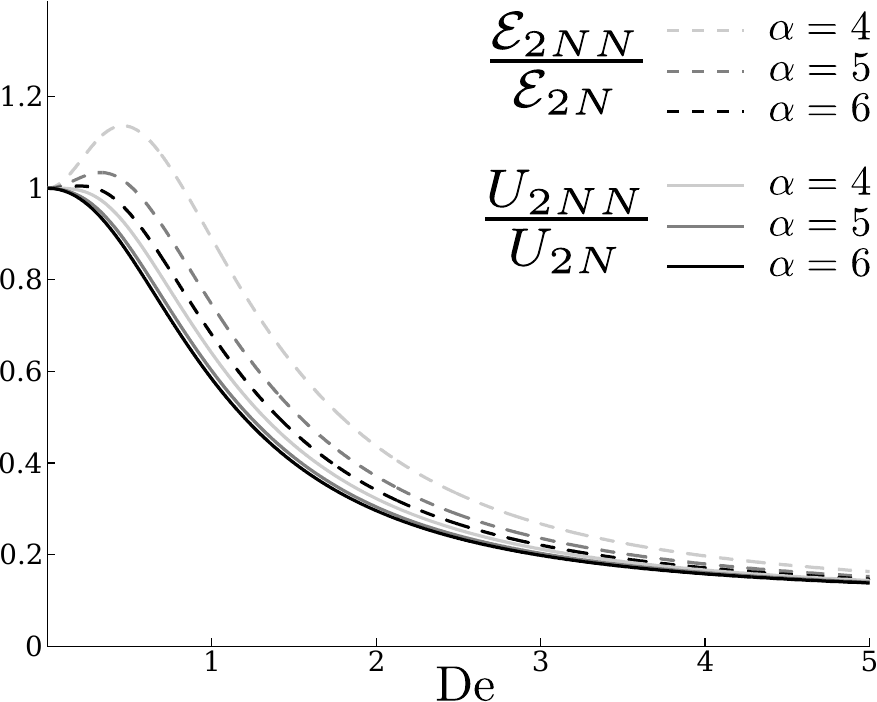}}
\caption{Three example waveforms are shown for which the swimming speed is not enhanced but the efficiency is. The relative amplitude in Eq.~\eqref{ss} lies outside the range  $n<\alpha<n^2$ ($n=2$).}
\label{effupspeeddown}
\end{figure}
\begin{equation}
\left(\frac{U_{2NN}}{U_{2N}}\right)^2 >  \frac{\eta_N W_{2NN}}{\eta W_{2N}}\cdot\label{49}
\end{equation}
\begin{figure*}[t]
\includegraphics[width=0.7\textwidth]{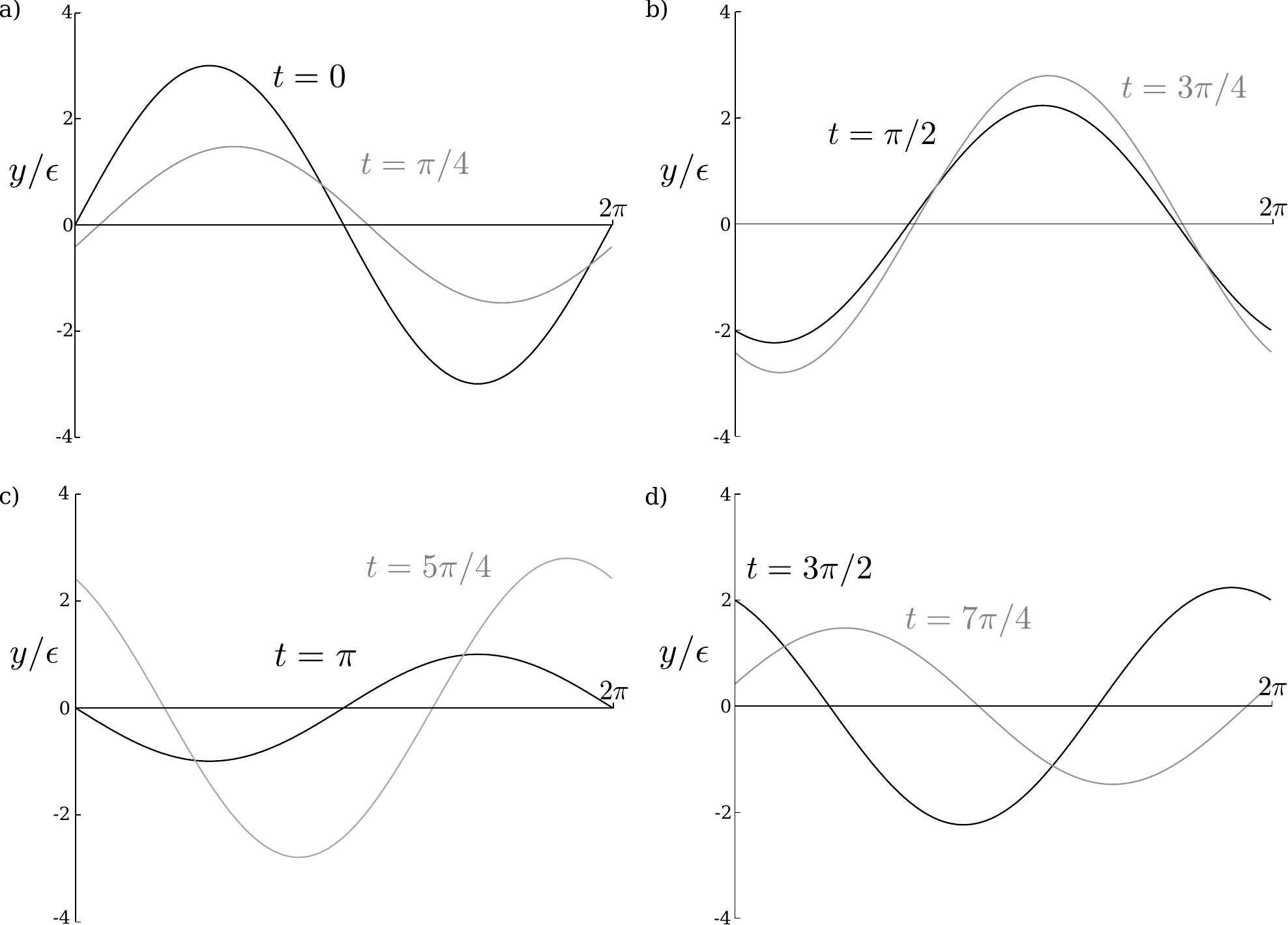}
\caption{Illustration of a waveform from Eq.~\eqref{sw} with $\alpha=2$ and $n=2$ that produces faster swimming in a non-Newtonian fluid. The black lines in each of the four figures show the waveform at 
dimensionless times  $0$, $\pi/2$, $\pi$ and $3\pi/2$, respectively,  and the grey lines show the evolution of the wave an eighth of a period later.}
\label{snaps2}
\end{figure*}
Given that the right-hand side of Eq.~\eqref{49} is less than one, the condition for enhanced efficiency does not require enhanced swimming. Specifically, using the illustrative sinusoidal waveform of Eq.~\eqref{ss}, we obtain an improved efficiency when  
\begin{equation}
\frac{ \displaystyle(n^3+\alpha^2)}{(\alpha^2-n^2)^2} > 
\displaystyle\frac
{\displaystyle\left[\alpha^2\left(\frac{1+\beta\De^2}{1+\De^2}\right)
+n^3\left(\frac{1+\beta n^2\De^2}{1+n^2\De^2}\right)\right]}
{\displaystyle\left[\alpha^2\left(\frac{1+\beta\De^2}{1+\De^2}\right)
-n^2\left(\frac{1+\beta n^2\De^2}{1+n^2\De^2}\right)\right]^2},
\end{equation}
for which $n<\alpha<n^2$ is not a necessary condition. This result  is illustrated in  Fig.~\ref{effupspeeddown} in the case $n=2$. When $\alpha>n^2$, the waveform travels in the same direction in both fluids and the swimmer is  always faster in a Newtonian fluid  although it is more efficient in the non-Newtonian one for a range of Deborah numbers.

\subsection{Two waves with identical wavelengths}
\setcounter{eqn}{0}

Instead of two waves with identical wave speeds,  enhanced swimming can also be obtained in a combination of waves with identical wavelengths. Since the waves need to have different frequencies, then they necessarily have different wave speeds. As an example  we consider here the waveform
\begin{equation}\label{sw}
 y=\epsilon[\alpha\sin(x-t)+\sin(x+nt)].
\end{equation}
This gives
\begin{equation}
 U_{2NN} = \frac{\alpha^2}{2}\left(\frac{1+\beta\De^2}{1+\De^2}\right)-\frac{n}{2}\left(\frac{1+n^2\De^2}{1+n^2\De^2}\right),
\end{equation} 
and
\begin{equation}
 U_{2N}=\frac{1}{2}(\alpha^2-n).
\end{equation}
Similarly as above, the second derivate of $U_{2NN}/U_{2N}$ at $\De=0$ is given by
\begin{equation}
 \frac{\partial^2}{\partial\De^2}\left(\frac{U_{2NN}}{U_{2N}}\right)\bigg|_{\De=0} = \left(\frac{\alpha^2-n^3}{\alpha^2-n}\right)(\beta-1),
\end{equation}
and   faster swimming occurs when
\begin{equation}
 n^{1/2}<\alpha<n^{3/2}, 
\end{equation}
which is confirmed by numerical computations (not shown). A waveform leading to enhanced swimming in this case is illustrated in Fig.~\ref{snaps2}, in the case $\alpha=2$ and $n=2$. 
This corresponds to a maximum speed enhancement of $U_{2NN}/U_{2N}\approx 1.1$ at Deborah number $\De\approx0.4$.
To obtain the optimal Deborah number, we extremise the ratio of swimming speeds to find the peaks occurring at 
\begin{equation}
 \De_{1\ast}=\sqrt{\frac{n^2-\alpha\sqrt{n}}{n^2(\alpha\sqrt{n}-1)}} \,\,\text{and}\,\, \De_{2\ast}=0,
\end{equation} 
with a behavior qualitatively similar to that of the last section. 

%%%%%%%%%%%
\section{Superposition of three travelling  waves: continuous vs.~discrete enhancement}
\setcounter{eqn}{0}
\begin{figure*}[t]
\begin{tabular}{cc}
  \putindeepbox[0pt]{\includegraphics[width=0.45\textwidth]{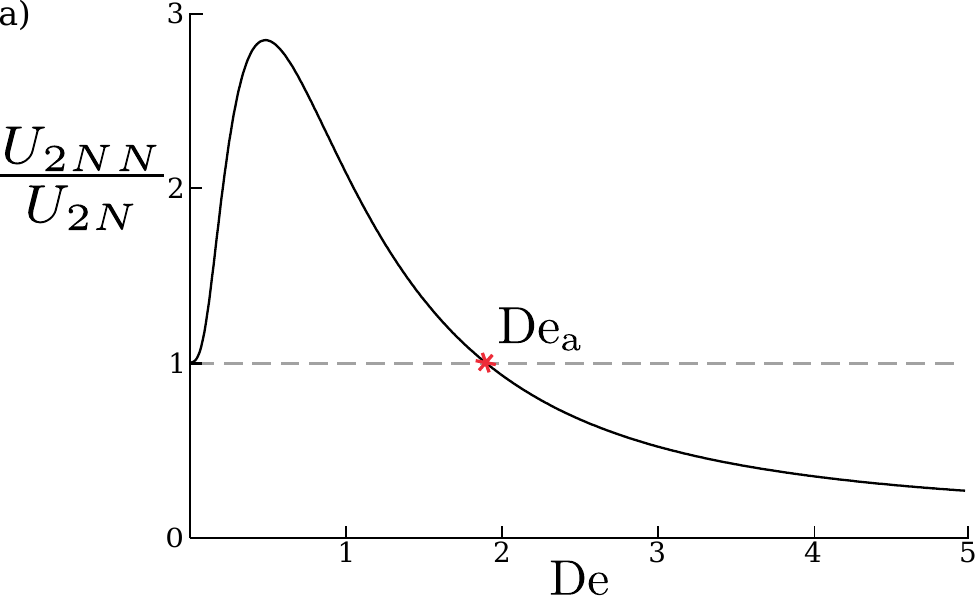}\label{figcon}}
    & \putindeepbox[0pt]{\includegraphics[width=0.45\textwidth]{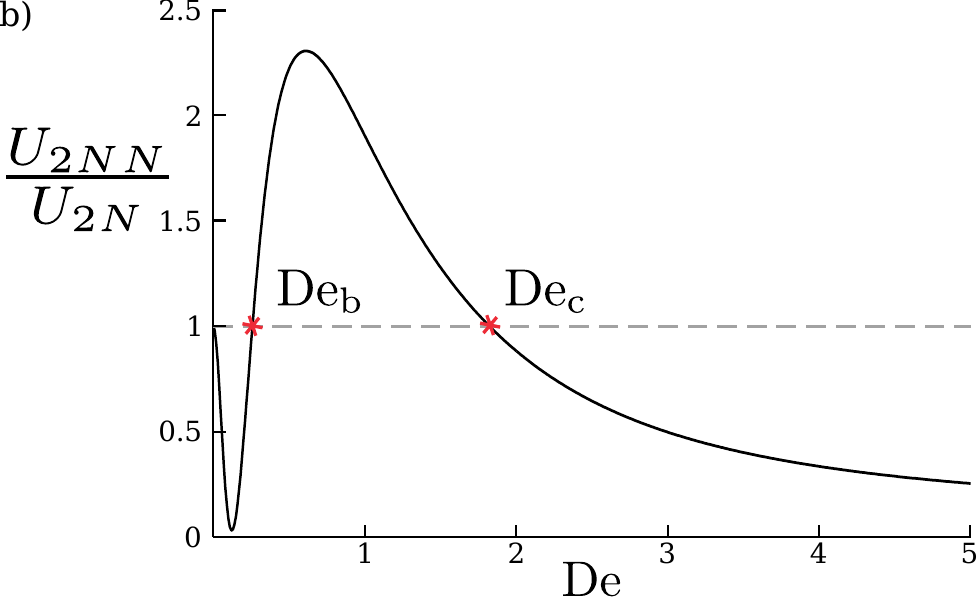}\label{fignoncon}}\\
\end{tabular}
\caption{Two different types of enhancement are shown for two different three-mode waves with $n_1=1$, $n_2=4$ and $n_3=8$: (a) continuous enhancement from zero Deborah number in the range $[0, \De_a]$ with $\De_a\approx1.9$ ($a_{n_1}= 1$, $a_{n_2}=-1$, $a_{n_3}=0.3$); 
(b) enhancement in a   discrete, finite,  range of  Deborah numbers, $[\De_b, \De_c]$, where $\De_b\approx0.3$ and $\De_c\approx1.8$ ($a_{n_1}=1$, $a_{n_2}=-2$, $a_{n_3}=1.2$).}
\label{figtypes}
\end{figure*}

In the previous sections, where the superposition of two waves was considered, we saw that when  enhancement is present, it is  continuous from $\De=0$ to an order one Deborah number $\De_a$, where the value $\De_a\neq0$ is the only non-zero solution of  $U_{2NN}/U_{2N}=1$. We now demonstrate that if the swimmer is able to use a third travelling  wave, it is possible for swimming enhancement to  occur only when  a finite amount of viscoelasticity is present, i.e.~for values of the Deborah number in the range $[\De_b, \De_c]$, where $\De_b$ and $\De_c$ 
are both non-zero.

We consider, for illustration purposes,  a waveform with three modes $1<n_2<n_3$. 
The corresponding square amplitudes $a_{1}$, $a_{n_2}$ and $a_{n_3}$ are non-dimensionalised by $a_1$ so that we take $a_1=1$. Using the same notation as above,  the Newtonian swimming speed is then given by
\begin{equation}\label{53}
 U_{2N} = 1+a_{n_2}+a_{n_3}.  
\end{equation}
The difference between the non-Newtonian and Newtonian swimming speeds is then found to be
\begin{align}\label{diffSS}
  U_{2NN}-U_{2N} = &\De^2(\beta-1)\Bigg[\left(\frac{1}{1+\De^2}\right)\notag\\
&+\left(\frac{n_2^2a_{n_2}}{1+n_2^2\De^2}\right)+\left(\frac{n_3^2a_{n_3}}{1+n_3^2\De^2}\right)\Bigg].
\end{align}
Focusing on  cases where $U_{2N}>0$, enhanced forward swimming is found when Eq.~\eqref{diffSS} is positive. As shown in Fig.~\ref{figtypes} numerically for $n_2=4$ and $n_3=8$, there are two types of enhancements possible: either on a range  $[0, \De_a]$ where the velocity ratio curves upward at the origin  (continuous enhancement, Fig.~\ref{figtypes}a, as in \S~\ref{twwvs}) or on a range  $[\De_b, \De_c]$ for which the curvature at $\De=0$ is initially negative  before curving upward as the viscoelasticity increases (discrete enhancement, Fig.~\ref{figtypes}b).

\begin{figure*}[t]
\includegraphics[width=0.85\textwidth]{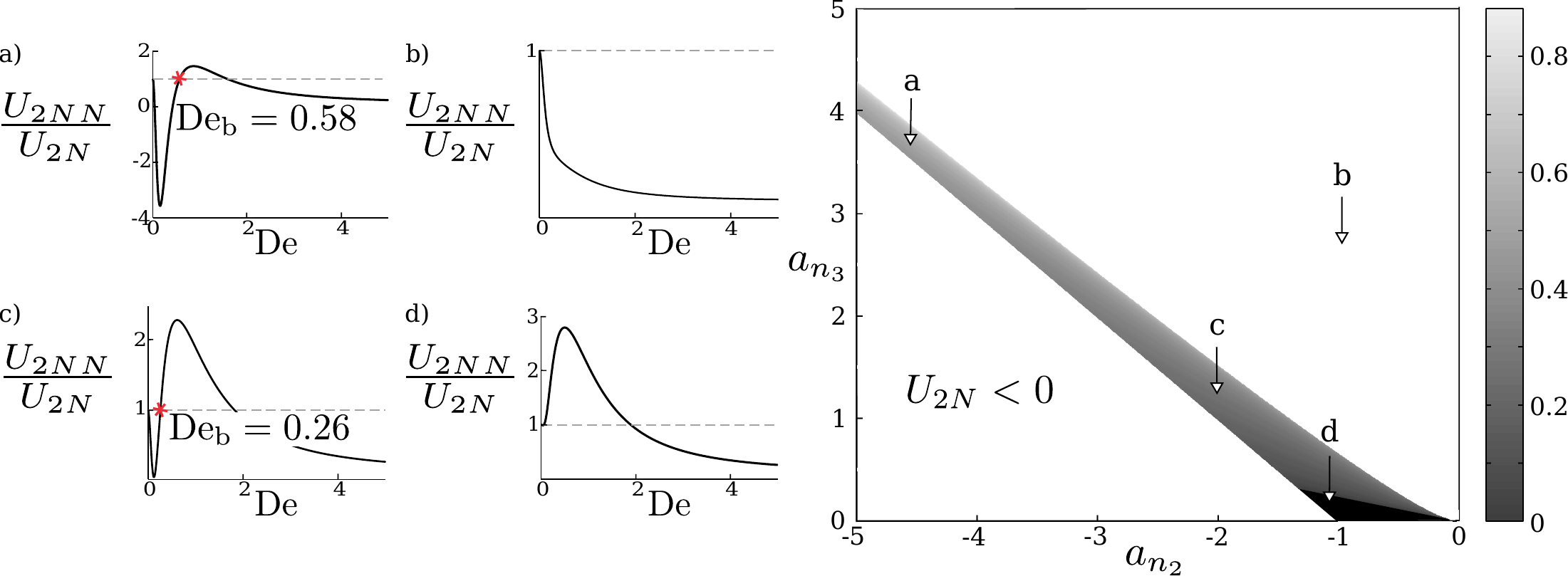}
\caption{Regions in the parameters space \{$a_{n_2},a_{n_3}$\}   where enhanced swimming  occurs. 
The black section represents where the curvature is positive (\emph{i.e.}~upwards curving) so that from infinitesimally small Deborah number we get  
an increase in the swimming speed. In contrast, the grey scale section shows regions where enhanced swimming is obtained despite negative curvature at the origin. 
The grey scale color scheme  quantifies the value of non-zero Deborah numbers at which the  increase in swimming speed first occurs, $\De_b$, from low Deborah number in dark, to high Deborah number ($\approx 0.8$) in light grey.
Results are shown for $n_2=4$ and $n_3=8$.}
\label{figblocks}
\end{figure*}

In order to distinguish between them analytically, we observe that when  the curvature 
is negative, we can either have no enhancement or enhancement at a finite Deborah number. Hence, we need to search for  cases where Eq.~\eqref{diffSS} is positive, given that the curvature at the origin  is negative. 
The curvature of the general wave was obtained  in Eq.~\eqref{curv}, hence for negative curvature in our three-mode waveform we require
\begin{equation}\label{cond2}
\kappa =  2(\beta-1)(1+n_2^2a_{n_2}+n_3^2a_{n_3})<0.
\end{equation}
The result in Eq.~\eqref{diffSS} can then be written in terms of $U_{2N}$ and $\kappa$ as
\begin{align}\label{cond3}
n_2^2n_3^2\De^4(\beta-1)&\left(\frac{n_2^2+n_3^2-1}{n_2^2n_3^2} +a_{n_2}+a_{n_3}\right)>\notag\\
&\left[\De^2\kappa(\tfrac12+\De^2)+n_2^2n_3^2\De^4(\beta-1)U_{2N}\right].
\end{align}
As $\beta-1<0$, and assuming that $U_{2N}>0$ and $\kappa<0$, the minimum requirement for  non-continuous enhancement is
\begin{equation}
\label{condsmin}
\frac{n_2^2+n_3^2-1}{n_2^2n_3^2}+a_{n_2}+a_{n_3}<0.
\end{equation}

The three conditions given by Eqs.~\eqref{53} ($U_{2N}>0$), \eqref{cond2}, and \eqref{condsmin} can  be satisfied simultaneously only when $a_{n_2}<0$ and $a_{n_3}>0$, \emph{i.e.}~the first and third modes  must have a different sign to the second mode. 
 We then search numerically  over the domain \{$a_{n_2}<0$, $a_{n_3}>0$\} and $\De$ to obtain the regions  where Eq.~\eqref{diffSS} is positive provided $U_{2N}>0$ and $\kappa<0$, in the {example} case $n_2=4$ and $n_3=8$. 
The values of $a_{n_2}$ and $a_{n_3}$ fitting these conditions are shown in Fig.~\ref{figblocks} (grey scale domain) while the region showing  continuous enhancement is shown in black. 
The grey scale colouring scheme used in Fig~\ref{figblocks} displays the value of the lower bound in the interval, $\De_b$, from low  (dark grey) to high (light grey) values. 
For three waves, in contrast to the case of two waves, situations  exist therefore where a finite amount of viscoelasticity is required to get enhanced propulsion, $\De > \De_b > 0$ 
(analysis for four and five mode waves show similar results and are not shown here).

%%%%%%%%%%%
\section{Discussion and conclusion}

Motivated by the non-Newtonian environment in which many swimmers propel themselves \emph{in vivo}, in this paper, we have calculated the speed of Taylor's  swimming sheet in a Newtonian and an Oldroyd-B (non-Newtonian) fluid, in the small-amplitude limit.  
In contrast  to previous analytical  studies, we found that small-amplitude travelling  waves can produce faster swimming in a non-Newtonian fluid compared to a Newtonian  fluid  when there are waves travelling  in opposite directions in different frequency modes and with different amplitudes. 
Physically, in a non-Newtonian fluid the waves in higher frequency modes are damped more than those in lower frequency modes, 
increasing the overall speed of the wave under conditions placed upon the difference in frequency and amplitude of the summed waveforms. 
The efficiency of the wave can also be increased, and the direction of swimming can sometimes be reversed.  
By studying in detail the superposition of two or three travelling  waves, we also showed that the range of Deborah number in which the enhancement of the swimming speed takes place can either include the origin, 
in which case any small amount of viscoelasticity will lead to faster swimming, or it may be a finite interval which does not include the origin, meaning that  faster locomotion requires a finite amount of viscoelasticity. 

The  results in this paper are reminiscent of recent experimental and theoretical work on the role of inertia in locomotion, where  two important questions have  been addressed: 
(1) for a non-swimmer at zero Reynolds numbers, how much inertia is needed to make it swim? and (2) how does the locomotion speed of a Stokesian swimmer vary with inertia? The answer to question 
(1) depends crucially on the geometry and actuation of the swimmer and both discrete \cite{Alben2004,Vandenberghe2004} and continuous \cite{Lauga2007P} transition to swimming were obtained. 
In response to question (2), model organisms called  squirmers were shown to vary their speed monotonically from zero Reynolds number~\cite{Wang2012,khair2014expansions}.  
Similarly, in our results, we showed that a careful design of the swimming kinematics could lead to either a decrease or an increase, which could be  continuous or discrete, of the locomotion speed.  
We  expect that these results will remain valid for more realistic models of swimming organisms, in particular those  including features such as large-amplitude, finite-size, and three-dimensional effects.

{In three-dimensions, the same frequency-dependent damping term as the one in Eq.~\eqref{final} is present for infinite cylindrical swimmers \cite{FuWolgemuthPowers2009}, hence similar results are expected to hold. With regards to finite sized swimmers, backward propagating  waves are expected to occur due to the finite nature of real flagella.  Previous computational studies have shown that the addition of viscoelasticity decreases the backwards motion of a finite swimmer \cite{Shelley2010} due to the presence of a viscoelastic network behind the swimmer.  The opposite has been observed  experimentally for nematodes where hyperbolic stresses created along the swimmer hinder propulsion \cite{Arratia2011}. It is yet unclear how our theoretical results would  extend to a finite swimmer,  though we may expect that the mechanism provided in this paper would provide an additional contribution to the swimming speed.  In the case of multiple swimmers, it would be interesting to investigate how   waves with both high and low frequency modes  affect one another and potentially synchronise. From Ref \cite{Elfring2010} we expect the synchronisation rate to increase with the frequency of the waves however the generalisation to multiple modes  in viscoelastic fluids has yet to be done.  A recent study address a related issue in the  Newtonian case \cite{Brumley2014}.}

The waveforms produced here offer insight into how swimming speeds can be increased in fluids with viscoelastic properties   often found in nature~\cite{RheoBook}. Can these shape kinematics  occur in biology? For flagellar swimming this requires understanding of how the stochastic actuation of molecular motors create waveforms.  Dynein, the motor protein causing flagella bending, has been proposed to have two distinct modes to create oscillatory bending - these can be described as active and passive, or forward and reverse active modes~\cite{Brokaw2009}, leading to  travelling  waves that can propagate up or down the flagellum. Due to the finite nature of flagella the wave is reflected back off the tail end or basal body, thus creating passive backwards waves~\cite{Machin1958}.
Experiments on \emph{Drosophila} spermatozoa show that the cells use actively created forwards and backwards flagellar waves to avoid obstacles~\cite{Yang2011}.  Furthermore by solving elastohydrodynamic force balance equations on infinite flagella analytic studies have shown two different modes of waves travel along the flagella with the same frequency, but different amplitudes and directions~\cite{Wiggins1998}.   Hence a flagellum naturally creates forward and backward travelling  waves with different amplitudes. {The    enhancement described in this paper 
requires however  waves with different frequencies  travelling in the backwards direction for enhancement to occur. 
The addition of higher frequency modes ($n=2$ and $n=3$), found in small amounts in beating spermatozoa \cite{ReidelKruse2007}, would not however lead to an increase in swimming speed as $a_n>0$ for all values of $n$ found experimentally.}  Changes in flagella beating frequency can occur by altering the environment in which the swimmer propagates,  for example hyperactivation when mammalian spermatozoa reach the ovum, leading to a reduction in  the beat frequency and increase in the beat amplitude~\cite{Suarez1991}; a variation in ATP or salt concentrations also change frequencies~\cite{Brokaw2009}. {Recent work on the unsteady modes of flagellar motion show that most  modes have a frequency smaller than the fundamental frequency, and  hence would correspond to a reduced swimming speed \cite{Bayly2015}. The addition of noise to the molecular motor oscillations, either through variations in concentrations in the bulk or variations between motors,  could lead to increased swimming provided the coherent noise is large enough for the flagellum to access a higher frequency mode, however this is  larger than the noise measured \cite{Ma2014}.}  Backwards travelling wave results have been described for muscle-actuated planar motion occurring for example in the nematode \emph{Caenorhabditis elegans}~\cite{Sznitman2010} as well as  other flagellar systems such as the green alga \emph{Chlamydomonas reinhardtii}~\cite{Guasto2010}. 

 While our study offers only an idealised view, {and although as of yet there are no  experimental studies from biology for which the model here  would predict faster swimming, our work address the most general periodic waving deformation and}  points to the  use of multiple waves travelling in different directions as a  mechanism allowing control of swimming magnitude and direction in complex environments.

\section*{Acknowledgements}
This work was funded in part by the European Union through a Marie Curie CIG grant to E.L.

\bibliographystyle{unsrt}
\bibliography{references}

\end{document}